\documentstyle[12pt,bezier]{article}
\newfont{\mathea}{msam10 scaled\magstep0}
\newfont{\matheb}{msbm10 scaled 1095}
\newfont{\tmpEins}{cmsy10 scaled 2074}
\newfont{\tmpZwei}{cmsy10 scaled 1095}
\newfont{\tmpDrei}{cmsy10 scaled 1000}
\newfont{\tmpVier}{cmsy5 scaled 1000}
\newfont{\tmpFuenf}{msbm7 scaled\magstep0}

\def\Bbb#1{\mathchoice{\mbox{\matheb #1}}{\mbox{\matheb #1}}%
 {\mbox{\tmpFuenf #1}}{\mbox{\tmpFuenf #1}}}

\def\restriction{\mathchoice{%diplaystyle
 \mbox{\unitlength1cm\begin{picture}(.2,.4)%
  \bezier{5}(.07,.3)(.1,.27)(.13,.24)%
  \put(.07,.35){\line(0,-1){.5}}\end{picture}}}{%textstyle
 \mbox{\unitlength1cm\begin{picture}(.2,.4)%
  \bezier{5}(.07,.3)(.1,.27)(.13,.24)%
  \put(.07,.35){\line(0,-1){.5}}\end{picture}}}{%scriptstyle
 \mbox{\mathea\symbol{22}}}{%scriptscriptstyle
 \mbox{\mathea\symbol{22}}}}

\def\dach#1#2{\mbox{$\mathop{\vbox{\ialign{%
  $##\crcr\hfil #1 \hfil$\crcr}}}\limits^{\scriptscriptstyle #2}$}}

\def\rnzs{\dach{\rho_2}{\mbox{$\scriptscriptstyle\kern-.7mm0$}}\kern-1.2mm'}

\def\Subset{\mbox{$\subset\kern-.5mm\subset$}}
\newcommand{\LI}{\mbox{{\rm L$^{\kern-.15em\raise.2ex\hbox{\scriptsize 1}}$}}}

\def\Ldummy{\left.\bgroup}
\def\Rdummy{\egroup^{\rule{0mm}{1.4mm}}\right.}
\def\LA{\left\langle\bgroup}
\def\RA{\egroup^{\rule{0mm}{1.4mm}}\right\rangle_{\cal A}^{}}
\def\LR{\left(\bgroup}
\def\RR{\egroup^{\rule{0mm}{1.4mm}}\right)}
\def\LG{\left\{\bgroup}
\def\RG{\egroup^{\rule{0mm}{1.4mm}}\right\}}

\def\Wort#1{\mbox{{\rm #1\kern.1em}}}

\def\lfac#1#2{\vcenter{\hbox{$#1\kern-.2em\raise-.6ex\hbox{\Large{/}}%
 \kern-.2em\raise-1.2ex\hbox{$#2$}$}}}

\def\gin{\mbox{\tmpZwei\symbol{91}\kern-1.4mm\rule{.2mm}{1.85mm}\kern1.4mm}}
\def\gni{\mbox{\tmpZwei\symbol{92}\kern-1.4mm\rule[.15mm]{.2mm}{1.85mm}%
  \kern1.4mm}}

\def\EINS{{\mathop{1\kern-.25em\mbox{{\rm{\small l}}}}}}

\begin{document}

\Large The eigenvalue problem for the resonances of the infinite-dimensional
Friedrichs model  on the positive half line with Hilbert-Schmidt perturbations

\vspace{1cm}

\normalsize Hellmut Baumg\"artel

\vspace{3mm}

Mathematical Institute, University of Potsdam

Am Neuen Palais 10, PF 601553

D-14415 Potsdam, Germany

e-mail: baumg@rz.uni-potsdam.de

\begin{abstract}

A Gelfand triplet for the Hamiltonian $H$ of the infinite-dimensional Friedrichs
model on the positive half line 
with Hilbert-Schmidt perturbations  is constructed
such that exactly the resonances (poles of the inverse of the Liv\v{s}ic-matrix) 
are  eigenvalues of 
the extension $H^{\times}$ of
$H$. The corresponding eigenantilinear forms are 
calculated explicitly. Using the wave matrices for the Abelian wave (M\"oller) operators 
the corresponding eigenantilinear forms for the unperturbed Hamiltonian $H_{0}$
turn out to be of pure Dirac type and can be characterized by their corresponding
Gamov vector which is uniquely determined by restriction to the intersection of the
Gelfand space 
for $H_{0}$
with $P_{+}{\cal H}^{2}_{+}$, where ${\cal H}^{2}_{+}$ is the 
Hardy space of the upper half plane. Simultaneously, this restriction yields a truncation
of the unitary evolution 
$t\rightarrow e^{-itH_{0}}$
to the well-known decay semigroup for $t\geq 0$ of the
Toeplitz type on $P_{+}{\cal H}^{2}_{+}$. That is, exactly those eigenvectors
$\lambda\rightarrow k(\lambda-\zeta)^{-1},\,,k$ element of the multiplicity space
${\cal K}$, of the
decay semigroup have an extension to an eigenantilinear form for $H_{0}$ hence for $H$
if $\zeta$ is a resonance and $k$ is from that subspace of ${\cal K}$ which is uniquely
determined by its corresponding Dirac type antilinear form. Moreover, the
scattering matrix which is meromorphic in the lower half plane has only simple
poles there and the main part of its Laurent representation is a linear combination
of Gamov vectors.

\end{abstract}

\vspace{3mm}

Keywords Resonances, Friedrichs model, scattering theory, Gamov vectors

\vspace{3mm}


\section{Introduction}

Breit-Wigner formulas $E\rightarrow c((E-E_{0})^{2}+(\Gamma/2)^{2})^{-1}$
describe bumps in quantum scattering cross sections, where $E_{0}$ is the
{\em resonance energy}, $\Gamma/2$ the {\em half width}. These bumps are 
associated to unstable particles with finite lifetimes. If the scattering matrix
is meromorphically continuable into the lower half plane, they can be connected with poles
$E_{0}-i(\Gamma/2)$ there. Then $c(E-E_{0})-i(\gamma/2))^{-1}$ is called the Breit-Wigner
amplitude (see e.g. Bohm [1]). These poles are called {\em resonances}. Their rigorous
mathematical description requires knowledge on the analytical properties of the
scattering matrix in dependence of the Hamiltonian $H$, which is difficult to 
obtain, in general.

A first step in this direction is to investigate the resolvent of $H$, the meromorphic
continuations of its matrix elements and their poles
which are candidates for resonances. These poles are then associated
with the
eigenvalues of non-selfadjoint operators connected with $H$. 
That is, already in this
approach
the emphasis is to obtain certain spectral properties of these poles.
The so-called
Aigular-Balslev-Combes-Simon theory is representative for this efforts (see
Aigular/Combes [2], Simon [3], see also Hislop/Sigal [4]). 

Another approach is to derive states associated with resonances which represent directly
their connection with decay, i.e. by satisfying the exponential decay law. These vectors
are called {\em Gamov vectors} in the literature (see e.g. Bohm/Gadella [5]).
The most simple method in this direction is to use the so-called decay semigroup
for $t\geq 0$ of the Toeplitz type (see Eisenberg et al [6], Strauss [7]) as a truncation
of the quantum evolution and to inspect its eigenvalue spectrum (which consists
of all points of the lower half plane). For this approach the Hardy spaces play an 
essential role. However, the crucial problem is then the selection of the  true Gamov
vectors $\lambda\rightarrow k/(\lambda-\zeta)^{-1}$. This requires explicit
knowledge of the scattering matrix, in particular for the calculation of the actual
parameters $k\in{\cal K}$, the multiplicity space.

The present paper presents a direct approach in the sense
to answer the question of the spectral properties of the resonances w.r.t. the Hamiltonian
$H$
for an infinite-dimensional Friedrichs-like
model (perturbation by Hibert-Schmidt operators). In this model the resonances
are characterized directly as the eigenvalues of an extension of $H$ by an appropriate 
Gelfand triplet. (Note that also for the "first step" triplets are used, e.g.
Banach space triplets by Agmon [8]). Moreover, from the eigenantilinear forms of the 
resonances, which act on the Gelfand space of the triplet, the associated Gamov vectors
(eigenvectors of the Toeplitz type semigroup) are uniquely derived by their 
restriction to the dense "Hardy space part" of the Gelfand space. Furthermore,
the analytic structure of the scattering matrix 
in the lower half plane
is characterized completely:
there are only simple poles there (among them are the resonances)
and the main part (in the Laurent sense) is a sum of a finite linear combination of
Gamov vectors (associated to the resonances) and a second part, associated to poles
which are not resonances (in this model).

The paper is related to [9]. There the finite-dimensional Friedrichs model on the whole
real line is considered within the Schwartz space framework. In this case one has
the full spectral analogy to the Lax-Phillips scattering theory. This fact 
suggested to combine 
Lax-Phillips ideas (mainly the Lax-Phillips semigroup) with the eigenvalue
problem for resonances in general.

\section{Preliminaries}
\subsection{Basic concepts of the model }
\subsubsection{Assumptions}

The scattering model considered is a modified infinite-dimensional Friedrichs model
on the positive half line where the perturbation is of the Hilbert-Schmidt type.
Put
\[
{\cal H}_{0,+}:=L^{2}(\Bbb{R}_{+},{\cal K}, d\lambda),\quad \Bbb{R}_{+}:=(0,\infty),
\]
where ${\cal K}$ is a separable multiplicity Hilbert space, and
\[
{\cal H}:={\cal H}_{0,+}\oplus{\cal E},
\]
where ${\cal E}$ is a separable Hilbert space, too. $H_{0}$ denotes the multiplication
operator on ${\cal H}_{0,+}$,
\[
H_{0}f(\lambda):=\lambda f(\lambda),\quad f\in{\cal H}_{0,+},
\]
and $A$ a selfadjoint compact operator on ${\cal E}$ with only positive eigenvalues.
Further let $\Gamma\in{\cal L}_{2}({\cal H})$ be a Hilbert-Schmidt operator
on ${\cal H}$ with $\Gamma f=0$ for $f\in{\cal H}_{0,+}$ and
$\Gamma {\cal E}\subseteq {\cal H}_{0,+}$, i.e.
\[
\Gamma e(\lambda)=M(\lambda)e,\quad e\in{\cal E},
\]
where $M(\lambda)$ is necessarily a Hilbert-Schmidt operator of
${\cal L}_{2}({\cal E}\rightarrow{\cal K})$, well-defined a.e. on $\Bbb{R}_{+},$
satisfying $\int_{0}^{\infty}\Vert M(\lambda)\Vert^{2}_{2}d\lambda<\infty$
because of
\[
\Vert\Gamma\Vert^{2}_{2}=\int_{0}^{\infty}\Vert M(\lambda)\Vert^{2}_{2}d\lambda.
\]
The perturbation is given by $\Gamma +\Gamma^{\ast}$, i.e. we put
\[
H:=(H_{0}\oplus A)+\Gamma+\Gamma^{\ast}.
\]
$H$ is selfadjoint and $\mbox{dom}\,H=\mbox{dom}\,H_{0}\oplus{\cal E}$.

In the finite-dimensional case where $\dim{\cal E}<\infty,\,\dim{\cal K}<\infty$ this
includes the case of the usual (finite-dimensional) Friedrichs model, where
$\Gamma$ is a partial isometry with $\Gamma^{\ast}\Gamma=P_{\cal E}$, the 
projection onto ${\cal E}$, and $\Gamma\Gamma^{\ast}<\EINS-P_{\cal E}$, such that
$\int_{0}^{\infty}M(\lambda)^{\ast}M(\lambda)d\lambda =\EINS_{\cal E}.$ 

In the following also the case of "small perturbations" is of interest, where
$\Gamma$ is replaced by $\epsilon\Gamma,\,0<\epsilon\leq 1,\,\epsilon$ the
so-called coupling constant.

For convenience in the following $\Gamma\restriction{\cal E}$ is identified with
$\Gamma$ without confusion. On this model we impose the following assumptions:

\begin{itemize}
\item[(i)]
${\cal H}_{0,+}=\mbox{clo\,spa}\{E_{0}(\Delta)f,\,f\in\Gamma{\cal E}\}$ and
${\cal H}=\mbox{clo\,spa}\{E(\Delta)e,\,e\in{\cal E}\}$, where
$E_{0}(\cdot),\,E(\cdot)$ denote the spectral measures of $H_{0},\,H$, respectively.
This means: $\Gamma{\cal E}$ is generating for ${\cal H}_{0,+}$ w.r.t.
$H_{0}$ and ${\cal E}$ is generating for ${\cal H}$ w.r.t. $H$. (In the case
$\dim\,{\cal E}<\infty$ this implies $\dim{\cal E}=\dim{\cal K}$.) If
$\dim{\cal E}=\infty$ then necessarily $\dim{\cal K}=\infty$.
\item[(ii)]
$H$ has no real eigenvalues.
\item[(iii)]
$\int_{0}^{\infty}\lambda^{2}\Vert M(\lambda)\Vert^{2}_{2}d\lambda<\infty$.
This is equivalent with $\Gamma{\cal E}\subset\mbox{dom}\,H_{0}$.
\end{itemize}

The essential parameter of the model is the Hilbert-Schmidt-valued operator
function $M(\cdot)$ on the positive half line. For this parameter we require the following
further conditions of analytic continuability:

\begin{itemize}
\item[(iv)]
The operator function $M(\cdot)$ is analytically continuable into the
complex plane,
\[
\Bbb{C}\ni z\rightarrow M(z)\in{\cal L}_{2}({\cal E}\rightarrow{\cal K}),
\]
which is rational, with no simple poles and with no poles on the real line.
That is, $M(\cdot)$ is holomorphic on the real line.
The point $\infty$ is a holomorphic point of $M(\cdot)$ and $M(\infty)=0$.

The set of all poles of $M(\cdot)$ is finite, the operator function
$\Bbb{C}\ni z\rightarrow M(\overline{z})^{\ast}$
is rational, too. The poles of this function are complex-conjugated to the
poles of $M(\cdot)$. The set of all these poles is denoted by ${\cal P}$, it is
a finite set, symmetric w.r.t. complex conjugation.
\item[(v)]
$\mbox{ima}\,M(z)=:{\cal A}\subset{\cal K}$
is a dense set in ${\cal K}$
for all $z\not\in{\cal P}$
which is independent of $z$.
\item[(vi)]
$\ker\,M(z)=\{0\}$ for all $z\not\in{\cal P}$, i.e. $z\rightarrow M(z)^{-1}$
exists for $z\not\in{\cal P}$.
\item[(vii)]
$\Bbb{C}\ni z\rightarrow M(z)^{-1}$ is a holomorphic operator function of
type A (in the sense of Kato [10]) on ${\cal A}$.
\end{itemize}

The conditions (v)-(vii) imply that
$M(z)=M(z_{0})C(z)$,
where $z_{0}\not\in{\cal P}$ and $z\rightarrow C(z)\in{\cal L}({\cal E})$
is a rational operator function with poles in ${\cal P}$
such that $C(z)$ is bounded invertible for $z\not\in{\cal P}$ and
$z\rightarrow C(z)^{-1}$ is holomorphic at the points of ${\cal P}$, too.

A simple example of $M(\cdot)$ satisfying these conditions is given in the case
${\cal E}={\cal K}=\Bbb{C}$ by $M(z):=(z+i)^{-2}$. Then
$M(z)^{-1}=(z+i)^{2}.$

\subsubsection{The operator function $\Phi$}

We put
\[
\Phi(z):=\Gamma^{\ast}R_{0}(z)\Gamma=\int_{0}^{\infty}
\frac{M(\lambda)^{\ast}M(\lambda)}{z-\lambda}d\lambda,\quad z\in\Bbb{C}_{>0},
\]
where $\Bbb{C}_{>0}$ is the complex plane with cut $[0,\infty)$,
i.e. $\Bbb{C}_{>0}:=\{z:z\in\Bbb{C}\setminus[0,\infty)\}.\;\Phi(\cdot)$
is holomorphic on $\Bbb{C}_{>0},\,\Phi(z)\in{\cal L}_{1}({\cal E}),\,
M(\lambda)^{\ast}M(\lambda)\in{\cal L}_{1}({\cal E})$ and
\[
\frac{\Gamma^{\ast}E_{0}(d\lambda)\Gamma}{d\lambda}=M(\lambda)^{\ast}M(\lambda),
\quad \lambda>0.
\]
By straightforward residual calculation one obtains
\begin{equation}
\Phi(z)=\log z\cdot M(\overline{z})^{\ast}M(z)-H(z),\quad z\in\Bbb{C}_{>0},
\end{equation}
where $\log(-1)=i\pi$ and $H(\cdot)$ denotes the (Laurent) main part of
$\log z\cdot M(\overline{z})^{\ast}M(z)$ 
in $\Bbb{C}_{>0}$.
That is, (1) confirms that
$\Phi(\cdot)$ is holomorphic on $\Bbb{C}_{>0}$ and shows that $\Phi(\cdot)$ is meromorphic
on the Riemannian surface of $z\rightarrow\log z.$ For convenience we use the
denotation $\Phi_{\pm}$ for $\Phi\restriction\Bbb{C}_{\pm}$ (where $\Bbb{C}_{\pm}$
are in the "first sheet" $\Bbb{C}_{>0}$. In the following $\Phi_{\pm}$
are mainly considered in $\Bbb{C}_{<0}$, the complex plane with cut
$(-\infty,0],$ i.e. $\Bbb{C}_{<0}:=\Bbb{C}\setminus(-\infty,0].$ Then
\[
\Phi_{-}(z)-\Phi_{+}(z)=2\pi iM(\overline{z})^{\ast}M(z),\quad z\in\Bbb{C}_{<0}.
\]
The poles of $\Phi_{+}$ are contained in $\Bbb{C}_{-}\cap{\cal P}$,
those of $\Phi_{-}$ are contained in $\Bbb{C}_{+}\cap{\cal P}$, i.e. there are only
finitely many poles. One has
\[
\sup_{z\in\Bbb{C}_{<0}(R)}
\Vert\Phi_{\pm}(z)\Vert<\infty,\quad \Bbb{C}_{<0}(R):=\Bbb{C}_{<0}\cap\{z:\vert z\vert
\geq R\},
\]
where $R$ is sufficiently large.

\subsubsection{Liv\v{s}ic-matrix and partial resolvent}

The Liv\v{s}ic-matrix is defined by
\[
L_{\pm}(z):=z\EINS_{\cal E}-A-\Phi_{\pm}(z),\quad z\in\Bbb{C}_{<0}.
\]
The function $z\rightarrow L_{\pm}(z)\in{\cal L}({\cal E})$ is meromorphic on
$\Bbb{C}_{<0}$, the poles are contained in $\Bbb{C}_{\mp}\cap{\cal P}.$

The sandwiched resolvent $R(z):=(z-H)^{-1}$ by the projection $P_{\cal E}$ is
called the {\em partial resolvent}. A straightforward calculation
(see e.g. [11, p. 136 f.]) gives
\begin{equation}
P_{\cal E}R(z)P_{\cal E}\restriction{\cal E}=L_{\pm}(z)^{-1},\quad z\in\Bbb{C}_{\pm}.
\end{equation}
(2) shows that $L_{\pm}(\cdot)^{-1}$ is holomorphic on $\Bbb{C}_{\pm}$. It is
meromorphic on $\Bbb{C}_{<0}$ because
\[
L_{\pm}(z)=z(\EINS_{\cal E}-z^{-1}(A+\Phi_{\pm}(z)))=z(\EINS_{\cal E}-K_{\pm}(z)),
\]
where $K_{\pm}(z):=z^{-1}(A+\Phi_{\pm}(z))$ is compact for all
$z\in\Bbb{C}_{<0}$. One has
$L_{\pm}(z)^{-1}=z^{-1}(\EINS_{\cal E}-K_{\pm}(z))^{-1}$.
Since
$z\rightarrow (\EINS_{\cal E}-K_{\pm}(z))^{-1}$
is holomorphic on $\Bbb{C}_{\pm}$, it follows that
$L_{\pm}(z)^{-1}$ is meromorphic on $\Bbb{C}_{<0}$, according to a well-known
result due to Keldysch [12]. A straightforward estimation gives
\begin{equation}
\sup_{z\in\Bbb{C}_{<0}(R)}\Vert L_{+}(z)^{-1}\Vert<\infty,\quad
\Bbb{C}_{<0}(R):=\Bbb{C}\cap\{z:\vert z\vert\geq R\},
\end{equation}
where $R$ is sufficiently large. Therefore $L_{+}(\cdot)^{-1}$ has at most
finitely many poles which are contained in $\Bbb{C}_{-}$. The set of these
poles is denoted by ${\cal R}$, they are called {\em resonances}. To simplify the
treatment it is assumed that the sets of poles of $L_{+}(\cdot)$ and
$L_{+}(\cdot)^{-1}$ are disjoint, ${\cal R}\cap{\cal P}=\emptyset$.

Applying Theorem 1 of [11, p. 139] to the partial resolvent one obtains:
$\lambda\rightarrow L_{\pm}(\lambda)^{-1}$
is holomorphic on $\Bbb{R}_{+}$ becauuse of (ii). 
Moreover, for $\lambda<0$ we have
$L_{+}(\lambda+i0)=L_{-}(\lambda-i0)$ and
$\lambda\rightarrow L_{\pm}(\lambda\pm i0)^{-1}$
are holomorphic there, again because of (ii).
Further we have

\vspace{3mm}

LEMMA 1. $\zeta_{0}\in\Bbb{C}_{-}$ {\em is a pole of} $L_{+}(\cdot)^{-1}$
{\em iff} $\dim\,\ker\,L_{+}(\zeta_{0})>0.$

\vspace{1mm}

Proof. If $\zeta_{0}$ is a pole of $L_{+}(\cdot)^{-1}$ then one has
$0\in\mbox{spec}\,L_{+}(\zeta_{0})=
\mbox{spec}\,\{\zeta_{0}(\EINS_{\cal E}-K_{+}(\zeta_{0}))\}$
and this means $0$ is an eigenvalue of
$\EINS_{\cal E}-K_{+}(\zeta_{0}),\,$ i.e. also an eigenvalue of
$L_{+}(\zeta_{0})$, hence
$\ker\,L_{+}(\zeta_{0})\supset\{0\}.$ The converse is obvious. $\Box$

\vspace{3mm}

The sandwiched spectral measure $E(\cdot)$ of $H$ satisfies
\begin{equation}
\frac{P_{\cal E}E(d\lambda)P_{\cal E}}{d\lambda}=
\frac{1}{2\pi i}
\left((L_{-}(\lambda)^{-1}-L_{+}(\lambda)^{-1}\right)=
L_{\pm}(\lambda)^{-1}M(\lambda)^{\ast}M(\lambda)L_{\mp}(\lambda)^{-1},
\quad \lambda>0.
\end{equation}
It vanishes for $\lambda<0.$ Therefore $\mbox{spec}\,H=[0,\infty)$
and it follows that $H$ is pure absolutely continuous.

\subsection{Spectral representations by spectral integrals}

The spectral integral
\[
{\cal H}_{0,+}\ni x:=\int_{0}^{\infty}E_{0}(d\lambda)\Gamma f(\lambda),\quad
\Bbb{R}_{+}\ni\lambda\rightarrow f(\lambda)\in{\cal E},
\]
exists iff $\int_{0}^{\infty}\Vert M(\lambda)f(\lambda)\Vert^{2}_{\cal K}d\lambda
<\infty.$
According to assumption (i) the linear manifold of all such spectral integrals
is dense in ${\cal H}_{0,+}$. It defines a spectral representation of
${\cal H}_{0,+}$ w.r.t. $H_{0}$ which is given by the isometric isomorphism between
${\cal H}_{0,+}$ and $L^{2}(\Bbb{R}_{+},\hat{\cal E}_{\lambda},d\lambda)$, where
the Hilbert space $\hat{\cal E}_{\lambda}$ is the completion of ${\cal E}$
w.r.t. the scalar product
$\langle e_{1},e_{2}\rangle_{\lambda}:=(M(\lambda)e_{1},M(\lambda)e_{2})_{\cal K}$.
Note that $\Gamma{\cal E}$ is a spectral manifold w.r.t. $E_{0}(\cdot)$. If
$\dim\,{\cal E}<\infty$ then $\hat{\cal E}_{\lambda}={\cal E}$ for all
$\lambda$. (See e.g. [13, p. 90 f.] for details.)

The same procedure for ${\cal H}$ and $H$ leads to a distinguished spectral
representation defined by the spectral integral
\begin{equation}
{\cal H}\ni y:=\int_{0}^{\infty}E(d\lambda)g(\lambda),\quad
\Bbb{R}_{+}\ni\lambda\rightarrow g(\lambda)\in{\cal E},
\end{equation}
where (5) exists iff
\[
\int_{0}^{\infty}\Vert M(\lambda)L_{\pm}(\lambda)^{-1}g(\lambda)\Vert
^{2}_{\cal K}d\lambda<\infty.
\]
That is, for these dense sets of spectral integrals we call $f(\cdot)$
the $E_{0}$-{\em representer} of $x,\,x(\lambda)=M(\lambda)f(\lambda)$, and
$g(\cdot)$ the $E$-{\em representer} of $y$. Note that if
$y\in\mbox{dom}\,H$ then $\lambda\rightarrow\lambda g(\lambda)$ is
the $E$-representer of $Hy$.

These two spectral representations we call the {\em natural} spectral
representations of ${\cal H},{\cal H}_{0}$ w.r.t. $H,H_{0}$, respectively.

\subsection{Wave operators, wave matrices and scattering matrix}

The strong wave operators are defined by
\begin{equation}
W_{\pm}:=\mbox{s-lim}_{t\rightarrow\pm\infty}e^{itH}e^{-itH_{0}}P^{0}_{ac},
\end{equation}
where $P^{0}_{ac}$ denotes the projection onto the absolutely continuous
subspace of $H_{0}$. If $H$ is pure absolutely continuous and the condition
of asymptotic completeness is satisfied then
\[
W_{\pm}^{\ast}=\mbox{s-lim}_{t\rightarrow\pm\infty}e^{itH_{0}}e^{-itH}.
\]
If the wave operators (6) exist, they can be rewritten as so-called
Abelian limits
\begin{equation}
W_{+}=\Omega_{+}:=\mbox{s-lim}_{\epsilon\rightarrow+0}
\int_{0}^{\infty}\epsilon e^{-\epsilon t}e^{itH}e^{-itH_{0}}P^{0}_{ac}dt,
\end{equation}
\begin{equation}
W_{-}=\Omega_{-}:=\mbox{s-lim}_{\epsilon\rightarrow+0}\int_{0}^{\infty}
\epsilon e^{-\epsilon t}e^{-itH}e^{itH_{0}}P^{0}_{ac}dt.
\end{equation}
Note that the existence of the Abelian limits $\Omega_{\pm}$ does not imply
the existence of the wave operators (6). The time integrals in (7) and (8)
can be transformed into spectral integrals (see [14, p. 361]):
\begin{equation}
\int_{0}^{\infty}\epsilon e^{-\epsilon t}e^{itH}e^{-itH_{0}}P_{ac}^{0}dt=
\int_{0}^{\infty}E(d\lambda)(\EINS_{\cal H}-VR_{0}(\lambda+i\epsilon))P_{ac}^{0},
\end{equation}
\begin{equation}
\int_{0}^{\infty}\epsilon e^{-\epsilon t}e^{-itH}e^{itH_{0}}P^{0}_{ac}dt=
\int_{0}^{\infty}E(d\lambda)(\EINS_{\cal H}-VR_{0}(\lambda-i\epsilon))P_{ac}^{0},
\end{equation}
where $V:=H-H_{0}=\Gamma+\Gamma^{\ast}$.

\vspace{3mm}

LEMMA 2. {\em The limits}
\begin{equation}
\Omega_{\pm}:=\mbox{s-lim}_{\epsilon\rightarrow+0}
\int_{0}^{\infty}E(d\lambda)(\EINS_{\cal H}-VR_{0}(\lambda\pm i\epsilon))P^{0}_{ac},
\end{equation}
\begin{equation}
\tilde{\Omega}_{\pm}:=\mbox{s-lim}_{\epsilon\rightarrow+0}
\int_{0}^{\infty}E_{0}(d\lambda)(\EINS_{\cal H}+VR(\lambda\pm i\epsilon))
\end{equation}
{\em exist}, $\Omega_{\pm}$ {\em is isometric from} ${\cal H}_{0,+}$ {\em onto}
${\cal H}$ {\em and} $\tilde{\Omega}_{\pm}=\Omega_{\pm}^{\ast}.$ {\em That is, the
Abelian limits of the wave operators exist and they are isometric
from} ${\cal H}_{0,+}$ {\em onto} ${\cal H}$.

\vspace{1mm}

Proof. The existence of (12) follows by straightforward calculation. The result is
\begin{equation}
\tilde{\Omega}_{\pm}y=\int_{0}^{\infty}E_{0}(d\lambda)\Gamma 
L_{\pm}(\lambda)^{-1}g(\lambda),
\end{equation}
where $y:=\int_{0}^{\infty}E(d\lambda)g(\lambda)$, i.e. $g(\cdot)$ is the
$E$-representer of $y$. (13) says that the $E_{0}$-representer of
$\tilde{\Omega}_{\pm}y$ is given by
$\lambda\rightarrow L_{\pm}(\lambda)^{-1}g(\lambda)$. One calculates easily
that
the "multiplication operator" $\lambda\rightarrow L_{\pm}(\lambda)^{-1}$
acts, w.r.t. the natural spectral representations of ${\cal H},{\cal H}_{0}$,
isometrically from ${\cal H}$ into ${\cal H}_{0}$, because
\[
\left\Vert\int_{0}^{\infty}E(d\lambda)g(\lambda)\right\Vert_{\cal H}^{2}=
\left\Vert\int_{0}^{\infty}E_{0}(d\lambda)\Gamma L_{\pm}(\lambda)^{-1}
g(\lambda)\right\Vert_{{\cal H}_{0,+}}^{2}.
\]
and the spectral integrals in ${\cal H}$ are dense, i.e.
\[\Vert\tilde{\Omega}_{\pm}y\Vert=\Vert y\Vert,\quad y\in{\cal H}
\]
follows. This means $\tilde{\Omega}_{\pm}$ is isometric on ${\cal H}$.
To show that the image of $\tilde{\Omega}_{\pm}$ is ${\cal H}_{0,+}$
it is sufficient to show that the dense set of all spectral integrals 
belongs to the image: choose an $E_{0}$-representer $\lambda\rightarrow f(\lambda)
\in{\cal E}$ of a vector $x\in{\cal H}_{0,+}$. Then
$\lambda\rightarrow L_{\pm}(\lambda)f(\lambda)\in{\cal E}$
is an $E$-representer of a vector $y\in{\cal H}$ because
\[
\left\Vert\int_{0}^{\infty}E(d\lambda)L_{\pm}(\lambda)f(\lambda)\right\Vert^{2}=
\int_{0}^{\infty}\left\Vert M(\lambda)f(\lambda)\right\Vert^{2}_{\cal K}
d\lambda<\infty,
\]
and one obtains $\tilde{\Omega}_{\pm}y=x$, hence
$\tilde{\Omega}_{\pm}{\cal H}={\cal H}_{0,+}$ follows or
$\tilde{\Omega}_{\pm}^{\ast}\tilde{\Omega}_{\pm}=\EINS_{\cal H},\;
\tilde{\Omega}_{\pm}\tilde{\Omega}_{\pm}^{\ast}=\EINS_{{\cal H}_{0,+}}$.
W.r.t. the representers of $x,y$ the operator
$\tilde{\Omega}_{\pm}^{\ast}$ is given by the transformation of
$\lambda\rightarrow f(\lambda)$ to
$\lambda\rightarrow L_{\pm}(\lambda)f(\lambda)$ or
\[
\tilde{\Omega}_{\pm}^{\ast}x=\int_{0}^{\infty}E(d\lambda)L_{\pm}(\lambda)f(\lambda).
\]
However, this can be rewritten into
\[
\tilde{\Omega}^{\ast}_{\pm}x=\int_{0}^{\infty}E(d\lambda)(\EINS_{\cal H}
-VR_{0}(\lambda\pm i0))x,\quad x\in{\cal H}_{0,+},
\]
i.e. (11) exists and $\Omega_{\pm}=\tilde{\Omega}^{\ast}_{\pm}.\quad \Box$

\vspace{3mm}

The proof of Lemma 2 yields

\vspace{3mm}

COROLLARY 1. {\em The (isometric) Abelian wave operators} $\Omega_{\pm},\,
\Omega_{\pm}^{\ast}$ {\em are given by the formulas}
\[\Omega_{\pm}\left(
\int_{0}^{\infty}E_{0}(d\lambda)\Gamma f(\lambda)\right)=
\int_{0}^{\infty}E(d\lambda)L_{\pm}(\lambda)f(\lambda),
\]
\[
\Omega_{\pm}^{\ast}\left(
\int_{0}^{\infty}E(d\lambda)g(\lambda)\right)=
\int_{0}^{\infty}E_{0}(d\lambda)\Gamma L_{\pm}(\lambda)^{-1}g(\lambda),
\]
{\em i.e. if}
$\lambda\rightarrow f(\lambda)$
{\em is the}
$E_{0}$-{\em representer of}
$x\in{\cal H}_{0,+}$
{\em then the}
$E$-{\em representer of}
$\Omega_{\pm}x\in{\cal H}$
{\em is given by}
$\lambda\rightarrow L_{\pm}(\lambda)f(\lambda).$
{\em Conversely, if}
$\lambda\rightarrow g(\lambda)$
{\em is the}
$E$-{\em representer of}
$y\in{\cal H}$
{\em then the }
$E_{0}$-{\em representer of}
$\Omega_{\pm}^{\ast}y\in{\cal H}_{0,+}$
{\em is given by}
$\lambda\rightarrow L_{\pm}(\lambda)^{-1}g(\lambda)$.

\vspace{3mm}

For example, the $E_{0}$-representer of the vector $\Omega_{+}^{\ast}e,\,e\in{\cal E}$,
is given by
\begin{equation}
\lambda\rightarrow M(\lambda)L_{+}(\lambda)^{-1}e,
\end{equation}
the $E_{0}$-representer of $\Omega_{+}^{\ast}\Gamma e$ is
\begin{equation}
\lambda\rightarrow M(\lambda)L_{+}(\lambda)^{-1}(\lambda-A)e,
\end{equation}
because $(\lambda-A)e$ is the $E$-representer of $\Gamma e$:
\[
\int_{0}^{\infty}E(d\lambda)(\lambda-A)e=
\int_{0}^{\infty}E(d\lambda)e-\int_{0}^{\infty}E(d\lambda)Ae=He-Ae=\Gamma e,
\]
note that $e\in\mbox{dom}\,H$, because $He=Ae+\Gamma e$.

Corollary 1 means that the Abelian wave operators act by application
of the Liv\v{s}sic-matrix resp. its inverse on the corresponding representers. 
In general, operator functions with these properties are called
{\em wave matrices} of $\Omega_{\pm},\,\Omega_{\pm}^{\ast}$. Note that wave
matrices are well-defined only if the spectral representations are fixed.

\vspace{3mm}

COROLLARY 2. {\em The wave matrices of} $\Omega_{\pm},\,\Omega_{\pm}^{\ast}$
{\em w.r.t. the natural spectral representations of}
${\cal H}_{0,+}\,{\cal H}$
{\em are given by}
$\Omega_{\pm}(\lambda)=L_{\pm}(\lambda),\;\Omega_{\pm}(\lambda)^{\ast}=
L_{\pm}(\lambda)^{-1},\,\lambda\in\Bbb{R}_{+}$.

\vspace{3mm}

Note that Lemma 2 implies that $H$ and $H_{0}$ are unitarily equivalent.

\vspace{3mm}

REMARK 1. Since $\Omega_{\pm}$ is isometric, (7) and (8) can be improved.
In this case even
\begin{equation}
\lim_{\epsilon\rightarrow+0}\int_{0}^{\infty}\epsilon e^{-\epsilon t}
\Vert e^{itH}e^{-itH_{0}}x-\Omega_{+}x\Vert^{2}dt=0,\quad
x\in{\cal H}_{0,+},
\end{equation}
\begin{equation}
\lim_{\epsilon\rightarrow+0}\int_{0}^{\infty}\epsilon e^{-\epsilon t}
\Vert e^{-itH}e^{itH_{0}}x-\Omega_{-}x\Vert^{2}dt=0, \quad
x\in{\cal H}_{0,+},
\end{equation}
are true. The conditions (16)\,,\,(17) are strongly related to the existence
of the strong limits (6), i.e. to the existence of the strong wave operators
(for details see [13, p. 101 f.]). If $\Gamma$ is even trace class,
$\Gamma\in{\cal L}_{1}({\cal H})$, then the strong limits (6) exist. In particular,
in the
finite-dimensional case $\dim\,{\cal E}<\infty$ the wave operators
$W_{\pm}$ exist.

The scattering operator $S$ is defined by $S:=W_{+}^{\ast}W_{-}=
\Omega_{+}^{\ast}\Omega_{-}$, it is unitary on
${\cal H}_{0,+}$ and can be represented by its {\em scattering matrix}
w.r.t. a given spectral representation of ${\cal H}_{0,+}$.

\vspace{3mm}

LEMMA 3. {\em W.r.t. the natural spectral representation of}
${\cal H}_{0,+}$
{\em the corresponding scattering matrix} $S_{\cal E}$
{\em is given by}
\[
S_{\cal E}(\lambda)=L_{+}(\lambda)^{-1}L_{-}(\lambda)=L_{+}(\lambda)^{-1}L_{+}(\lambda)^{\ast},
\]
{\em i.e. if}
$x\in{\cal H}_{0,+}$
{\em and}
$f(\cdot)$
{\em is its}
$E_{0}${\em -representer then}
$\lambda\rightarrow S_{\cal E}(\lambda)f(\lambda)$
{\em is the}
$E_{0}${\em -representer of}
$Sx.\quad S_{\cal E}(\cdot)$
{\em is meromorphically continuable into}
$\Bbb{C}_{<0}$, {\em its poles are contained in}
${\cal R}\cup({\cal P}\cap\Bbb{C}_{+})$.

\vspace{3mm}

The scattering matrix $S_{\cal K}(\cdot)$ w.r.t. the original spectral
representation of ${\cal H}_{0,+}$ given by the ${\cal K}$-valued functions
$\lambda\rightarrow x(\lambda)\in{\cal K}$ for $x\in{\cal H}_{0,+}$
satisfies
\[
S_{\cal K}(\lambda)M(\lambda)f(\lambda)=M(\lambda)S_{\cal E}(\lambda)f(\lambda)
\]
which is satisfied by
\begin{equation}
S_{\cal K}(\lambda):=\EINS_{\cal K}-2\pi i M(\lambda)L_{+}(\lambda)^{-1}M(\lambda)^{\ast}.
\end{equation}
$S_{\cal K}(\cdot)$ is meromorphically continuable into
$\Bbb{C}_{<0}$, its poles are contained in ${\cal R}\cup{\cal P}$.

\vspace{1mm}

In the case of "small perturbations" where $\Gamma$ is replaced by $\epsilon\Gamma$,
i.e. $M(\cdot)$ is replaced by $\epsilon M(\cdot)$, there is an essential
difference between the resonances (poles of $L_{+}(\cdot)^{-1}$) and the poles
of $L_{-}(\cdot)$ (which are in $\Bbb{C}_{+}$) because the poles of
$z\rightarrow\epsilon M(z)$ and $z\rightarrow\epsilon M(\overline{z})^{\ast}$
are independent of $\epsilon$, whereas the poles of
$L_{+}(\cdot)^{-1}$ 
depend on $\epsilon$. 
This implies that also these poles of
$S_{\cal K}(\cdot)$, contained in ${\cal R}$, depend on $\epsilon$, those in
${\cal P}$ are independent of $\epsilon$. The most interesting resonances
are those whose trajectories for $\epsilon\rightarrow 0$ run into the embedded
eigenvalues, i.e. the eigenvalues of $A$.

\section{The Gelfand triplet}

A Gelfand triplet is given by the Gelfand space and its topology.
 
\vspace{3mm}

The {\em Gelfand space} ${\cal G}\subset{\cal H}_{0,+}$ is defined to be the
manifold of all $s\in{\cal H}_{0,+}$,
\[
s(\lambda)=M(\lambda)L_{+}(\lambda)^{-1}g(\lambda),\quad g(\lambda)\in{\cal E},
\]
($g(\cdot)$ is the $E$-representer of $g:=\Omega_{+}s$), such that $g(\cdot)$
is holomorphic on $\Bbb{R}_{+}$, meromorphic on $\Bbb{C}_{<0}$ with poles
at most in ${\cal P}$, and $\lambda\rightarrow\lambda g(\lambda)$ is also
an $E$-representer.
The {\em Gelfand topology} in ${\cal G}$ is defined by the collection of norms
\[
{\cal G}\ni s\rightarrow[s]_{\cal K}:=\Vert s\Vert_{{\cal H}_{0,+}}+
\sup_{z\in K}\Vert g(z)\Vert_{\cal E},\quad K\subset\Bbb{C}_{<0}\setminus{\cal P},
\quad K\; \mbox{compact}.
\]
Obviously ${\cal G}$ is dense in ${\cal H}_{0,+}$ w.r.t. the Hilbert topology and
${\cal G}$ defines a Gelfand triplet
\[
{\cal G}\subset{\cal H}_{0,+}\subset{\cal G}^{\times},
\]
where ${\cal G}^{\times}$ denotes the set of all continuous antilinear forms w.r.t. the
Gelfand topology. Note that
\[
\Omega_{+}^{\ast}{\cal E}\oplus\Omega_{+}^{\ast}\Gamma{\cal E}\subset{\cal G}.
\]
This follows from (14) and (15). Note further that
$\Omega_{+}^{\ast}\Gamma{\cal E}\subset\mbox{dom}\,H_{0}$
because of (iii).

The Gelfand space ${\cal G}$ can be transferred into ${\cal H}$ by the wave operator
$\Omega_{+}$:
\[
{\cal D}:=\Omega_{+}{\cal G}.
\]
The topology of ${\cal D}$ is defined by the injection of the topology of
${\cal G}$. Then ${\cal D}$ is a Gelfand space in ${\cal H}$ and
\[
{\cal D}\subset{\cal H}\subset{\cal D}^{\times}
\]
is the corresponding Gelfand triplet. ${\cal D}$ satisfies
\[
{\cal E}\oplus\Gamma{\cal E}\subset{\cal D},\quad {\cal D}\subseteq\mbox{dom}\,H,
\quad H{\cal D}\subseteq{\cal D}.
\]
Then, defining $\Phi$ by $\Phi:=P_{\cal E}^{\bot}{\cal D}
\subset{\cal D}$ one obtains
\[
{\cal D}=\Phi\oplus{\cal E},\quad {\cal D}^{\times}=\Phi^{\times}\times{\cal E}
\quad (\mbox{cartesian product}).
\]
Further one has
$\Phi\subseteq\mbox{dom}\,H_{0}$ because of
$\Phi\subseteq\mbox{dom}\,H=\mbox{dom}\,H_{0}\oplus{\cal E}$, thus
$H_{0}\Phi\subseteq\Phi$ follows.

For ${\cal D}\ni d=\phi+e,\;\phi\in\Phi,\,e\in{\cal E}$ and
$d^{\times}=\{\Phi^{\times},e^{\times}\}\in\Phi^{\times}\times{\cal E}$
one obtains
\begin{equation}
\langle d\mid d^{\times}\rangle=
\langle\phi\mid\phi^{\times}\rangle+(e,e^{\times})_{\cal E}.
\end{equation}

\section{The eigenvalue problem for $H$ w.r.t. the Gelfand triplet
${\cal D}\subset{\cal H}\subset{\cal D}^{\times}$}
\subsection{The boundary condition and the solution}

The Gelfand triplet ${\cal D}\subset{\cal H}\subset{\cal D}^{\times}$
yields a unique extension $H^{\times}$ on ${\cal D}^{\times}$ given by
\begin{equation}
\langle d\mid H^{\times}d^{\times}\rangle :=\langle Hd\mid d^{\times}\rangle,
\quad d\in{\cal D},\,d^{\times}\in{\cal D}^{\times}.
\end{equation}
The eigenvalue equation for eigenvalues $\zeta_{0}\in\Bbb{C}_{-}\setminus{\cal P}$
of $H^{\times}$ reads then
\begin{equation}
H^{\times}d_{0}^{\times}=\zeta_{0}d_{0}^{\times},
\quad d_{0}^{\times}:=\{\phi_{0}^{\times}(\zeta_{0},e_{0}\},\quad
e_{0}\in{\cal E},\,\zeta_{0}\in\Bbb{C}_{-}\setminus{\cal P},\,
\phi_{0}^{\times}\in\Phi^{\times}.
\end{equation}
For the part $\phi_{0}^{\times}$ of a solution we impose a

\vspace{1mm}

{\em Boundary condition}: $\phi_{0}^{\times}$ is required to be the analytic
continuation into $\Bbb{C}_{-}$ across $\Bbb{R}_{+}$ of a holomorphic
vector antilinear form $\phi_{0}^{\times}(z,e_{0})$ on $\Bbb{C}_{+}$
such that the $\Phi$-part of the eigenvalue equation is an identity
on $\Bbb{C}_{+}$.

\vspace{1mm}

The solution of this eigenvalue problem is given by

\vspace{3mm}

THEOREM 1. {\em The point}
$\zeta_{0}\in\Bbb{C}_{-}\setminus{\cal P}$
{\em is an eigenvalue of}
$H^{\times}$
{\em with eigenantilinear form}
$d_{0}^{\times}:=\{\phi_{0}^{\times}(\zeta_{0},e_{0}),e_{0}\}$
{\em iff}
$\zeta_{0}$
{\em is a resonance,}
$\zeta_{0}\in{\cal R}$,
{\em and}
$e_{0}$
{\em satisfies}
$L_{+}(\zeta_{0})e_{0}=0,$
{\em i.e.}
$e_{0}\in\ker\,L_{+}(\zeta_{0}).$
{\em That is, the (generalized) eigenspace of}
$\zeta_{0}$
{\em has the dimension}
$\dim\ker\,L_{+}(\zeta_{0}),$
{\em the geometric multiplicity of the eigenvalue} $0$
{\em of}
$L_{+}(\zeta_{0}).$

\vspace{1mm}

Proof. According to (19) and (20) the eigenvalue equation (21) means
\[
\langle Hd\mid d_{0}^{\times}\rangle=\langle\overline{\zeta}_{0}d\mid d_{0}^{\times}
\rangle\quad d\in{\cal D},
\]
This is equivalent with
\[
(Ae-\overline{\zeta}_{0}e,e_{0})+\langle\Gamma e\mid\phi_{0}^{\times}\rangle=
\langle\overline{\zeta}_{0}\phi-H_{0}\phi\mid\phi_{0}^{\times}\rangle
-(\Gamma^{\ast}\phi,e_{0}),
\]
where $d=\phi+e,\,\phi\in\Phi,\,e\in{\cal E}$. Since $e$ and $\phi$ vary 
independently we obtain two equations:
\begin{equation}
((\overline{\zeta}_{0}-A)e,e_{0})=\langle\Gamma e\mid\phi_{0}^{\times}\rangle,
\quad e\in{\cal E},
\end{equation}
and
\begin{equation}
\langle (\overline{\zeta}_{0}-H_{0})\phi\mid\phi_{0}^{\times}\rangle=
(\Gamma^{\ast}\phi,e_{0}),\quad \phi\in\Phi.
\end{equation}
$\phi_{0}^{\times}$ depends on $\zeta_{0}$, the possible eigenvalue (and on $e_{0}$).
According to the boundary condition for $\phi_{0}^{\times}$
this antilinear form is required to be the analytic continuation of
a holomorphic vector antilinear form
$\Bbb{C}_{+}\ni z\rightarrow\phi_{0}^{\times}(z)$
such that the equation (23) is valid also on $\Bbb{C}_{+}$:
\begin{equation}
((\overline{z}-H_{0})\phi,\phi_{0}^{\times}(z))_{{\cal H}_{0,+}}=
(\Gamma^{\ast}\phi,e_{0})_{\cal E},\quad z\in\Bbb{C}_{+},\,\phi\in\Phi,
\end{equation}
or
\[
(\phi,(z-H_{0})\phi_{0}^{\times}(z))_{{\cal H}_{0,+}}=
(\phi,\Gamma e_{0})_{{\cal H}_{0,+}},\quad z\in\Bbb{C}_{+},\,\phi\in\Phi.
\]
This means $(z-H_{0})\phi_{0}^{\times}(z)=\Gamma e_{0}$ or
\[
\phi_{0}^{\times}(z,e_{0})=(z-H_{0})^{-1}\Gamma e_{0},\quad z\in\Bbb{C}_{+}.
\]
That is, with $\phi:=P_{\cal E}^{\bot}\Omega_{+}s,\;s\in{\cal G}$, one obtains
\[
\langle\phi\mid\phi_{0}^{\times}(z,e_{0})\rangle=
(P_{\cal E}^{\bot}\Omega_{+}s,R_{0}(z)\Gamma e_{0})_{{\cal H}_{0,+}}=
(\Omega_{+}s,R_{0}(z)\Gamma e_{0})_{{\cal H}_{0,+}},\quad z\in\Bbb{C}_{+}.
\]
Now one has to check that this antilinear form has an analytic continuation into
$\Bbb{C}_{-}$ across $\Bbb{R}_{+}$ which is meromorphic on $\Bbb{C}_{<0}.$
First we use the following identity which can be obtained by a straightforward
calculation:
\[
R_{0}(z)\Gamma e+e=R(z)L_{+}(z)e,\quad e\in{\cal E},\,z\in\Bbb{C}_{+}.
\]
Then we get
\[
(\Omega_{+}s,R_{0}(z)\Gamma e_{0})=
(\Omega_{+}s,R(z)L_{+}(z)e_{0})-(\Omega_{+}s,e_{0}).
\]
$(\Omega_{+}s,e_{0})$ is a constant term. We put $g:=\Omega_{+}s\in{\cal D}$
and obtain
\begin{eqnarray}
(g,R(z)L_{+}(z)e_{0}) &=&
(R(\overline{z})g,L_{+}(z)e_{0})\nonumber \\
&=& \left(
\int_{0}^{\infty}\frac{1}{\overline{z}-\mu}E(d\mu)g(\mu),L_{+}(z)e_{0}\right)\nonumber \\
&=& \left(
\int_{0}^{\infty}\frac{1}{\overline{z}-\mu}\frac{P_{\cal E}E(d\mu)P_{\cal E}}
{d\mu}g(\mu)d\mu,L_{+}(z)e_{0}\right) \nonumber\\
&=& \left(
\int_{0}^{\infty}\frac{1}{\overline{z}-\mu}L_{-}(\mu)^{-1}M(\mu)^{\ast}
M(\mu)L_{+}(\mu)^{-1}g(\mu)d\mu,L_{+}(z)e_{0}\right) \nonumber\\
&=& (\Psi_{-}(\overline{z}),L_{+}(z)e_{0}),
\end{eqnarray}
where
\[
\Psi_{\pm}(z):=\int_{0}^{\infty}\frac{1}{z-\mu}L_{-}(\mu)^{-1}M(\mu)^{\ast}
M(\mu)L_{+}\mu)^{-1}g(\mu)d\mu,\quad z\in\Bbb{C}_{\pm},
\]
$\Psi_{\pm}(\cdot)$ is meromorphically continuable into $\Bbb{C}_{<0}$
across $\Bbb{R}_{+}$, where
\[
\Psi_{-}(z)-\Psi_{+}(z)=
2\pi i L_{-}(z)^{-1}M(\overline{z})^{\ast}M(z)L_{+}(z)^{-1}g(z),\quad z\in\Bbb{C}_{<0},
\]
and for $z\in\Bbb{C}_{-}$ one obtains
\[
\Psi_{-}(\overline{z})=\Psi_{+}(\overline{z})+2\pi i L_{-}(\overline{z})^{-1}
M(z)^{\ast}M(\overline{z})L_{+}(\overline{z})^{-1}g(\overline{z})
\]
and
\begin{eqnarray}
(\Psi_{-}(\overline{z}),L_{+}(z)e_{0}) &=&
(\Psi_{+}(\overline{z}),L_{+}(z)e_{0})+2\pi i(M(\overline{z})L_{+}(\overline{z})^{-1}
g(\overline{z}),M(z)e_{0}) \nonumber\\
&=&(\Psi_{+}(\overline{z}),L_{+}(z)e_{0})+2\pi i(L_{+}(\overline{z})^{-1}g(\overline{z}),
M(\overline{z})^{\ast}M(z)e_{0}).
\end{eqnarray}
Inspection of (26) proves the assertion. Poles are necessarily in ${\cal P}$.

Now we know that the antilinear form $\phi_{0}^{\times}(z,e_{0})$
satisfies the equation (24) for $z\in\Bbb{C}_{+}$. Therefore it satisfies
the equation (23) for all $z\in\Bbb{C}_{<0}\setminus{\cal P}$
and it is holomorphic there. For this reason we consider the second equation
(22) first on $\Bbb{C}_{+}$. Then it reads
\[
((\overline{z}-A)e,e_{0})=\langle\Gamma e\mid\phi_{0}^{\times}(z,e_{0})\rangle=
(\Gamma e,R_{0}(z)\Gamma e_{0})=(e,\Gamma^{\ast}R_{0}(z)\Gamma e_{0}).
\]
Using $(z-A)e_{0}-\Gamma^{\ast}R_{0}(z)\Gamma e_{0}=L_{+}(z)e_{0}$ we have
\begin{equation}
(e,(z-A)e_{0})-\langle\Gamma e\mid\phi_{0}^{\times}(z,e_{0})\rangle=
(e,L_{+}(z)e_{0}),\quad e\in{\cal E}\,z\in\Bbb{C}_{+}
\end{equation}
and the equation (22) reads simply $(e,L_{+}(z)e_{0})=0$
for all $e\in{\cal E}$ which obviously has no solution in
$\Bbb{C}_{+}\cup\Bbb{R}_{+}$. But by analytic continuation the identity (27)
is true also in $\Bbb{C}_{-}$. That is, equation (22) is equivalent with
\begin{equation}
L_{+}(\zeta_{0})e_{0}=0,\quad \zeta_{0}\in\Bbb{C}_{-}\setminus{\cal P}.
\end{equation}
This means: equation (22) has a solution $\zeta_{0}\in\Bbb{C}_{-}\setminus{\cal P}$
with the corresponding parameter $e_{0}\in{\cal E}$ iff equation (28) is
satisfied. Conversely, if $\zeta_{0}\in\Bbb{C}_{-}\setminus{\cal P}$ and
$e_{0}\in{\cal E}$ satisfy equation (28) then $\zeta_{0}$ is an eigenvalue
of $H^{\times}$ and $d_{0}^{\times}:=\{\phi_{0}^{\times}(\zeta_{0},e_{0}),e_{0}\}$
is a corresponding eigenantilinear form. The dimension of the eigenspace
of $\zeta_{0}$ is then $\dim\ker\,L_{+}(\zeta_{0}).\quad \Box$

\subsection{Characterization of the eigenantilinear forms by the Gelfand
triplet ${\cal G}\subset{\cal H}_{0,+}\subset{\cal G}^{\times}$}

The solutions $d_{0}^{\times}(\zeta_{0},e_{0})$ of the eigenvalue problem
for $H^{\times}$ refer to the Gelfand triplet
${\cal D}\subset{\cal H}\subset{\cal D}^{\times}$ which is given by the transfer
${\cal D}=\Omega_{+}{\cal G}$ from the Gelfand triplet of the Hilbert space
${\cal H}_{0,+}$. The "back transformation"
\begin{equation}
s_{0}^{\times}(\zeta_{0},e_{0}):=(\Omega_{+}^{\ast})^{\times}d_{0}^{\times}
(\zeta_{0},e_{0}),\quad s_{0}^{\times}\in{\cal G}^{\times}
\end{equation}
of the eigensolution $d_{0}^{\times}$ to the Gelfand triplet
${\cal G}\subset{\cal H}_{0,+}\subset{\cal G}^{\times}$
has a surprising property.

\vspace{3mm}

THEOREM 2. {\em The eigenantilinear form}
$s_{0}^{\times}$
{\em of}
$H_{0}^{\times}$
{\em w.r.t. the Gelfand triplet}
${\cal G}\subset{\cal H}_{0,+}\subset{\cal G}^{\times},$
{\em associated to}
$d_{0}^{\times}$ {\em by} (29) {\em is of pure Dirac type w.r.t. the point}
$\overline{\zeta}_{0}\in\Bbb{C}_{+}$:
\[
{\cal G}\ni s\rightarrow\langle s\mid s_{0}^{\times}(\zeta_{0},e_{0})\rangle :=
2\pi i(s(\overline{\zeta}_{0}),k_{0})_{\cal K},
\]
{\em where} $k_{0}:=M(\zeta_{0})e_{0}$.

\vspace{1mm}

Proof. For $z\in\Bbb{C}_{+}$ one has
\begin{eqnarray*}
\langle s\mid s_{0}^{\times}(z,e_{0})\rangle &=&
\langle\Omega_{+}s\mid d_{0}^{\times}(z,e_{0})\rangle \\
&=& \langle P_{\cal E}^{\bot}\Omega_{+}s\mid\phi_{0}^{\times}(z,e_{0})\rangle+
(P_{\cal E}\Omega_{+}s,e_{0}),
\end{eqnarray*}
i.e.
\[
\langle s\mid s_{0}^{\times}(z,e_{0})\rangle=(\Omega_{+}s,R(z)L_{+}(z)e_{0}).
\]
For $z\in\Bbb{C}_{-}$ one has, according to (25) and (26),
\[
\langle s\mid s_{0}^{\times}(z,e_{0})\rangle=
(\Psi_{-}(\overline{z}),L_{+}(z)e_{0})+
2\pi i(M(\overline{z})L_{+}(\overline{z})^{-1}g(\overline{z}),M(z)e_{0}).
\]
If $z=\zeta_{0}$ is a resonance then
\[
\langle s\mid s_{0}^{\times}(\zeta_{0},e_{0})\rangle=
2\pi i(s(\overline{\zeta}_{0}),k_{0}),\quad k_{0}:=M(\zeta_{0})e_{0}
\]
and this is the assertion.$\quad \Box$

\subsection{The associated Gamov vectors}

According to Theorem 2, the back transformed eigenantiliner forms
$s_{0}^{\times}(\zeta_{0},e_{0})$
are of pure Dirac type. This property is crucial for the association of
Gamov vectors which are uniquely determined by $s_{0}^{\times}(\zeta_{0},e_{0})$.

In the literature there are several approaches to associate "Gamov vectors" to resonances.
In one of them Gamov vectors are considered to be special eigenvectors
of a {\em truncated evolution}
$t\rightarrow T_{+}(t),t\geq 0,$ on the Hilbert space
$P_{+}{\cal H}^{2}_{+}$, where
$T_{+}(t):=P_{+}Q_{+}e^{-it\tilde{H}_{0}}P_{+}^{-1}$
and $\tilde{H}_{0}$ denotes the extension of $H_{0}$ to the multiplication operator
on ${\cal H}_{0}:=L^{2}(\Bbb{R},{\cal K},d\lambda),\; P_{+}$
the projection of ${\cal H}_{0}$ onto ${\cal H}_{0,+},\, Q_{+}$
the projection of ${\cal H}_{0}$ onto the Hardy space ${\cal H}^{2}_{+}$
(see e.g. Eisenberg et al [6], Strauss [7], see also [15]). The truncated
evolution is a strongly continuous contraction semigroup on
$P_{+}{\cal H}^{2}_{+}$ of the Toeplitz type (see Strauss [7]). As it is
well-known, each point $\zeta\in\Bbb{C}_{-}$
is an eigenvalue of the generator of this semigroup and the corresponding
eigenspace is given by 
$\{P_{+}f: f\in{\cal H}^{2}_{+},f(\lambda):=k(\lambda-\zeta)^{-1},k\in{\cal K}\}$,
i.e. the dimension of the eigenspace of $\zeta$
coincides with $\dim\,{\cal K}$.

A crucial question is wether Gamov vectors of this type can be uniquely associated to 
eigenantilinear forms
of $H_{0}^{\times}$.
A first answer is that one has to select the poles of $L_{+}(\cdot)^{-1}$ resp.
of $S_{\cal K}(\cdot)$. However, it remains the question which values of
$k\in{\cal K}$ have to be chosen. To solve this problem we consider an
appropriate dense subset of ${\cal G}$ by means of the Hardy space
${\cal H}_{+}^{2}$.

\vspace{3mm}

LEMMA 4. {\em The inclusions}
\[
{\cal G}\cap P_{+}{\cal H}^{2}_{+}\subset P_{+}{\cal H}^{2}_{+}\subset{\cal H}_{0,+}
\]
{\em are dense inclusions w.r.t. the Hilbert topology of}
${\cal H}_{0,+}$.

\vspace{1mm}

Proof. The density of the inclusion
$P_{+}{\cal H}^{2}_{+}\subset{\cal H}_{0,+}$
is a standard result. To prove the density of the first inclusion
we consider all Schwartz functions
$\Bbb{R}_{+}\ni\lambda\rightarrow v(\lambda)\in{\cal E}$
with compact support. Then the functions
$\lambda\rightarrow M(\lambda)v(\lambda)$
are ${\cal K}$-valued Schwartz functions with compact support in $\Bbb{R}_{+}$.
The linear span of all these functions is denoted by
${\cal U}\subset{\cal H}_{0,+}.\;{\cal U}$
is dense in ${\cal H}_{0,+}$. Note that
the functions of ${\cal U}$ can be considered also as functions on $\Bbb{R}$, i.e.
as functions from ${\cal H}_{0}$ and that the functions
$\lambda\rightarrow(M(\lambda)v(\lambda))'$ belong to ${\cal H}_{0}$, too.
Then
\[
(F^{-1}u(z):=\int_{0}^{\infty}e^{iz\lambda}u(\lambda)d\lambda,\quad
z\in\Bbb{C},\,u\in{\cal U},
\]
is holomorphic on $\Bbb{C}$ and one has 
$F^{-1}u\in{\cal H}^{2}_{+}$ and $(F^{-1}u)(z)\in{\cal A}$.
Then
$w:=P_{+}F^{-1}u\in P_{+}{\cal H}^{2}_{+}$ and $w\in\mbox{dom}\,H_{0}$.
Moreover, $P_{+}F^{-1}{\cal U}$ is dense in $P_{+}{\cal H}^{2}_{+}$.
Put
\[
s(\lambda):=M(\lambda)^{-1}w(\lambda),\quad \lambda\in\Bbb{R}_{+}.
\]
Then $s(z):=M(z)^{-1}w(z)\in{\cal E}$
is holomorphic on $\Bbb{C}_{<0}$. Now we define $g(\cdot)$ by
\[
g(z):=L_{+}(z)s(z)\in{\cal E},\quad z\in\Bbb{C}_{<0}\setminus{\cal P}.
\]
Then $z\rightarrow g(z)$ is holomorphic on $\Bbb{C}_{<0}\setminus{\cal P}$
and
\[
w(z)=M(z)L_{+}(z)^{-1}g(z).
\]
This means that $g$ is the $E$-representer of $w$. Therefore one obtains
$w\in{\cal G}$ because also $\lambda\rightarrow\lambda g(\lambda)$
is an $E$-representer. Hence
$P_{+}F^{-1}{\cal U}\subset{\cal G}\cap P_{+}{\cal H}^{2}_{+}$ follows, i.e.
a fortiori also ${\cal G}\cap P_{+}{\cal H}^{2}_{+}$
is dense in $P_{+}{\cal H}^{2}_{+}.\quad \Box$

\vspace{3mm}

Now we introduce in ${\cal G}\cap P_{+}{\cal H}^{2}_{+}$ a third topology by
\[
P_{+}{\cal H}^{2}_{+}\ni f\rightarrow\langle f\rangle:=\Vert
P_{+}^{-1}f\Vert_{{\cal H}_{0}}.
\]
Note that the projection $P_{+}$ restricted to ${\cal H}^{2}_{+}$
with image $P_{+}{\cal H}^{2}_{+}$ is a bijection (see e.g [13]). Moreover,
$P_{+}{\cal H}^{2}_{+}$ is a Hilbert space w.r.t. the norm $\langle\cdot\rangle.$

\vspace{3mm}

LEMMA 5. {\em The inclusion}
${\cal G}\cap P_{+}{\cal H}^{2}_{+}\subset P_{+}{\cal H}^{2}_{+}$
{\em is also dense w.r.t. the Hilbert norm}
$\langle\cdot\rangle$
{\em of}
$P_{+}{\cal H}^{2}_{+}$.

\vspace{1mm}

Proof. It is obvious from the proof of Lemma 4 because
$P_{+}^{-1}({\cal G}\cap P_{+}{\cal H}^{2}_{+})\supset F^{-1}{\cal U}$
which is dense in ${\cal H}^{2}_{+}$ w.r.t. its Hilbert norm. $\quad \Box$

\vspace{3mm}

The next step to obtain associated Gamov vectors for the eigenantilinear forms
$s_{0}^{\times}(\zeta_{0},e_{0})$ is to restrict them from
${\cal G}$ to ${\cal G}\cap P_{+}{\cal H}^{2}_{+}$.

\vspace{3mm}

COROLLARY 3. {\em The restricted eigenantilinear form}
$s_{0}^{\times}(\zeta_{0},e_{0})\restriction{\cal G}\cap P_{+}{\cal H}^{2}_{+}$
\[
{\cal G}\cap P_{+}{\cal H}^{2}_{+}\ni s\rightarrow
2\pi i(s(\overline{\zeta}_{0}),k_{0})_{\cal K},\quad
k_{0}=M(\zeta_{0})e_{0},
\]
{\em is even continuous w.r.t. the Hilbert topology}
$\langle\cdot\rangle$ {\em of} $P_{+}{\cal H}^{2}_{+}$,
{\em i.e. it can be continuously extended to}
$\mbox{clo}_{\langle\cdot\rangle}({\cal G}\cap P_{+}{\cal H}^{2}_{+})=P_{+}{\cal H}^{2}
_{+}$.
{\em That is},
$s_{0}^{\times}(\zeta_{0},e_{0})\restriction P_{+}{\cal H}^{2}_{+}$
{\em is realized by the}
$P_{+}{\cal H}^{2}_{+}${\em vector}
\[
\Bbb{R}_{+}\ni\lambda\rightarrow\frac{k_{0}}{\zeta_{0}-\lambda}
\]
{\em via the relation}
\begin{equation}
2\pi i(s(\overline{\zeta}_{0}),k_{0})_{\cal K}=
\int_{-\infty}^{\infty}\left(s(\lambda),\frac{k_{0}}{\zeta_{0}-\lambda}\right)_{\cal K}
d\lambda,
\end{equation}
{\em where in} (30) {\em the (unique) extensions of} $s(\cdot)$ {\em and}
$\lambda\rightarrow k_{0}(\zeta_{0}-\lambda)^{-1}$
{\em onto the whole real line have to be used.}

\vspace{1mm}

Proof. It follows immediately from the Paley-Wiener theorem.$\quad\Box$

\vspace{3mm}

Corollary 3 means: the restriction to $P_{+}{\cal H}^{2}_{+}$ of the eigenantilinear form
$s_{0}^{\times}(\zeta_{0},e_{0})$,
associated to the resonance $\zeta_{0}$ and to the parameter vector
$e_{0}\in\ker L_{+}(\zeta_{0})$,
which is the back transform
$S_{0}^{\times}(\zeta_{0},e_{0})=(\Omega_{+}^{\ast})^{\times}
d_{0}^{\times}(\zeta_{0},e_{0})$ of $d_{0}^{\times}(\zeta_{0},e_{0})$
to the Hilbert space ${\cal H}_{0,+}$ resp. to the corresponding Gelfand triplet, 
yields the associated Gamov vector
$\lambda\rightarrow k_{0}(\zeta_{0}-\lambda)^{-1}$ where
$k_{0}=M(\zeta_{0})e_{0}$. Conversely, exactly those eigenvectors
$\lambda\rightarrow k(\zeta-\lambda)^{-1}$
of the Toeplitz type semigroup $T_{+}(\cdot)$
have an extension (or "continuation") to an eigenantilinear form of the extended 
Hamiltonian $H_{0}^{\times}$ resp. to the extended Hamiltonian $H^{\times}$
if $\zeta=\zeta_{0}$ is a resonance and $k=k_{0}=M(\zeta_{0})$ with
$e_{0}\in\ker L_{+}(\zeta_{0})$.
That is, Corollary 3 solves the problem of the selection of the "true"
Gamov vectors.

Obviously, the parameter spaces $M(\zeta_{0})\ker L_{+}(\zeta_{0})$
resp. $\ker L_{+}(\zeta_{0})$ can be expressed by the scattering
matrix $S_{\cal E}(\cdot)$ at $\zeta_{0}$. Moreover, if $\zeta_{0}$ is a simple
pole of $S_{\cal E}(\cdot)$ then the parameter space can be calculated
using the Laurent expansion of $S_{\cal E}(\cdot)$ at $\zeta_{0}$.
Note that $S_{\cal E}(z)^{-1}=L_{-}(z)^{-1}L_{+}(z)$ is holomorphic
in $\Bbb{C}_{-}\setminus{\cal P}$, hence at $\zeta_{0}$.

\vspace{3mm}

PROPOSITION 1. {\em One has} $\ker L_{+}(\zeta_{0})=\ker S_{\cal E}(\zeta_{0})^{-1}$.
{\em If} $\zeta_{0}$ {\em is a simple pole of}
$S_{\cal E}$ {\em then}
\begin{equation}
\ker L_{+}(\zeta_{0})=\mbox{ima}\,\{\mbox{Res}_{z=\zeta_{0}}S_{\cal E}(z)\}.
\end{equation}

\vspace{1mm}

Proof. An easy calculation gives
\[
\ker L_{+}(\zeta_{0})=\mbox{ima}\,L_{-1}=\mbox{ima}(L_{-1}L_{-}(\zeta_{0})),
\]
where $L_{-1}=\mbox{Res}_{z=\zeta_{0}}L_{+}(z)^{-1}$. This gives (31). Note
that $L_{-}(\zeta_{0})^{-1}$ exists and is bounded. $\quad \Box$

\vspace{3mm}

REMARK 3. $\zeta_{0}$ is a simple pole of $L_{+}(\cdot)^{-1}$, i.e. of
$S_{\cal E}(\cdot)$, iff $\zeta_{0}$ is a simple pole of $S_{\cal K}(\cdot)$
because $S_{-1}\neq 0$ iff $M(\zeta_{0})S_{-1}M(\overline{\zeta}_{0})^{\ast}\neq 0$,
where $S_{-1}$ denotes the residuum of $\zeta_{0}$ as a pole of $L_{+}(\cdot)^{-1}$.
This is obvious from (18) because $M(\zeta_{0})S_{-1}M(\overline{\zeta}_{0})^{\ast}
=0$ implies, according to (vi), $S_{-1}M(\overline{\zeta}_{0})^{\ast}=0$ or
$M(\overline{\zeta}_{0})S_{-1}^{\ast}=0$, hence $S_{-1}^{\ast}=0=S_{-1}$
follows. The converse is trivial.

Surprisingly it turns out that $S_{\cal K}(\cdot)$ has {\em only} simple poles
in $\Bbb{C}_{-}$. This is pointed out in the next section.

\section{The function-theoretic characterization of the scattering
matrix $S_{\cal K}(\cdot)$ on the lower half plane}

According to (18) the scattering matrix
$z\rightarrow S_{\cal K}(z)$
is meromorphic on $\Bbb{C}_{<0}$. In $\Bbb{C}_{-}$ there are poles at the resonances 
(points of ${\cal R}$), there are no other poles there because of
${\cal R}\cap{\cal P}=\emptyset$. Possible poles in $\Bbb{C}_{+}$ are at the points of
${\cal P}$. Since $S_{\cal K}(\cdot)$ is unitary on the positive half line one has
\begin{equation}
S_{\cal K}(z)^{-1}=S_{\cal K}(\overline{z})^{\ast},\quad z\in\Bbb{C}_{<0}.
\end{equation}
On the upper border of $\Bbb{C}_{<0}$ (the negative half line)
$S_{\cal K}(\cdot)$ is holomorphic, i.e.
$\lambda\rightarrow S_{\cal K}(\lambda+i0)$ is holomorphic for $\lambda<0$
because
$L_{+}(\lambda+i0)^{-1}=P_{\cal E}(\lambda-H)^{-1}P_{\cal E}\restriction{\cal E}$
for $\lambda<0$ and $(\lambda-H)^{-1}$ is holomorphic there.
Further we have
\[
L_{+}(\lambda-i0)=L_{-}(\lambda-i0)+2\pi iM(\lambda)^{\ast}M(\lambda),\quad \lambda<0.
\]
That is, $L_{+}(\lambda-i0)$ is holomorphic for $\lambda<0$, hence
$L_{+}(\lambda-i0)^{-1}$ remains meromorphic for $\lambda<0$ and there is no pole
on the negative half line: 

In the contrary case, if $-\lambda_{0},\,\lambda_{0}>0$,
is a pole then, according to Lemma 1, we have
$\ker L_{+}(-\lambda_{0}-i0)\supset\{0\}$, i.e. there is $e_{0}\in{\cal E}$
such that
\[
(L_{-}(-\lambda_{0}-i0)+2\pi i M(-\lambda_{0})^{\ast}M(-\lambda_{0}))e_{0}=0
\]
or
\[
(-\lambda_{0}-A+\int_{0}^{\infty}\frac{M(\mu)^{\ast}M(\mu)}{\lambda_{0}+\mu}d\mu
+2\pi i M(-\lambda_{0})^{\ast}M(-\lambda_{0}))e_{0}=0
\]
hence
\[
(e_{0},(\lambda_{0}+A)e_{0})=\left(e_{0},
\int_{0}^{\infty}\frac{M(\mu)^{\ast}M(\mu)}{\lambda_{0}+\mu}d\mu e_{0}\right)
+2\pi i\Vert M(-\lambda_{0})e_{0}\Vert_{\cal E}^{2}.
\]
This implies $M(-\lambda_{0})e_{0}=0$ and $e_{0}=0$.

Therefore, $S_{\cal K}(\cdot)$ is holomorphic also on the lower border of
$\Bbb{C}_{<0}$, i.e. $\lambda\rightarrow S_{\cal K}(\lambda-i0)$ is holomorphic
for $\lambda<0$. Then from (32)
\[
S_{\cal K}(\lambda\pm i0)^{-1}=S_{\cal K}(\lambda\mp i0)^{\ast}
\]
follows, i.e. $S_{\cal K}(\lambda\pm i0)$ is bounded invertible for
$\lambda<0$, but not necessarily unitary. Moreover, from (3) and (iv) we have
\[
\sup_{z\in\Bbb{C}_{<0}(R)}\Vert S_{\cal K}(z)\Vert_{\cal K}<\infty,\quad
\Bbb{C}_{<0}(R):=\Bbb{C}_{<0}\cap\{z:\vert z\vert\geq R\},
\]
where $R$ is sufficiently large. Now it turns out that these three conditions
\begin{itemize}
\item $z\rightarrow S_{\cal K}(z)$ is meromorphic on $\Bbb{C}_{<0}$,
holomorphic on $\Bbb{R}_{+}$,
\item there exist the norm limits $\lim_{\epsilon\rightarrow+0}
S_{\cal K}(\lambda\pm i\epsilon)=:S_{\cal K}(\lambda\pm i0)$ for $\lambda<0$,
holomorphic on $\Bbb{R}_{-}:=(-\infty,0)$,
\item $S_{\cal K}(\cdot)$ is bounded at infinity on $\Bbb{C}_{<0}$
\end{itemize}
are sufficient to describe the function-theoretic behaviour of $S_{\cal K}(\cdot)$
on the lower half plane (note that these conditions already imply that
there are at most finitely many poles in $\Bbb{C}_{-}$).

First we extend the scattering operator $S$, originally a unitary operator
on ${\cal H}_{0,+}$, to a bonded operator on ${\cal H}_{0}$ using the
boundary values $S_{\cal K}(\lambda\pm i0)$ for $\lambda<0$. We define
the bounded operator
\[
{\cal H}_{0}\ni f\rightarrow S_{\pm}f\in{\cal H}_{0}
\]
by
\[(S_{\pm}f)(\lambda):=\left\{
\begin{array}{ll}
S_{\cal K}(\lambda)f(\lambda), & \lambda>0,\\
S_{\cal K}(\lambda\pm i0)f(\lambda), & \lambda <0.
\end{array} \right.\]
Then it turns out that for vectors $g\in{\cal H}^{2}_{-}$ the projection
of $S_{-}g$ onto the Hardy space ${\cal H}^{2}_{+}$ has a very simple form:

\vspace{3mm}

THEOREM 3. {\em The relation}
\begin{equation}
(Q_{+}S_{-}g)(z)=\sum_{\zeta\in{\cal R}\cup{\cal P}_{-}}
\frac{S_{-1,\zeta}g(\zeta)}{z-\zeta},\quad z\in\Bbb{C}_{+},\,g\in{\cal H}^{2}_{-},
\end{equation}
{\em is valid, where} $S_{-1,\zeta}$ {\em denotes the residuum of}
$S_{\cal K}(\cdot)$ {\em at the pole} $\zeta$ {\em and}
${\cal P}_{-}:={\cal P}\cap\Bbb{C}_{-}$.

\vspace{1mm}

Proof. It is given in [15], see also Gadella [16] for related calculations.$\quad\Box$

\vspace{3mm}

Since the right hand side of (33) is also well-defined on $\Bbb{C}_{-}$ and
rational on $\Bbb{C}$ we have

\vspace{3mm}

COROLLARY 4. {\em The poles of the scattering matrix} $S_{\cal K}(\cdot)$ {\em in}
$\Bbb{C}_{-}$ {\em are necessarily simple, the Laurent representation of}
$S_{\cal K}(\cdot)$ {\em in} $\Bbb{C}_{-}$ {\em is given by}
\begin{equation}
S_{\cal K}(z)=\sum_{\zeta\in{\cal R}\cup{\cal P}_{-}}
\frac{S_{-1,\zeta}}{z-\zeta}+H_{\cal K}(z),\quad z\in\Bbb{C}_{-},
\end{equation}
{\em where} $z\rightarrow H_{\cal K}(z)$ {\em is holomorphic on} $\Bbb{C}_{-}$.

\vspace{1mm}

Proof. The projection of $S_{-}g$ onto ${\cal H}^{2}_{-}$
is given by
\[
(Q_{-}S_{-}g)(z)=-\frac{1}{2\pi i}\int_{-\infty}^{\infty}
\frac{S_{-}(\lambda)g(\lambda)}{\lambda-z}d\lambda,\quad z\in\Bbb{C}_{-}.
\]
The right hand side of (33) is a rational function on $\Bbb{C}$, i.e.
$(Q_{+}S_{-}g)(\cdot)$ has an analytic continuation into $\Bbb{C}_{-}$
given by this expression. On the other hand
$Q_{+}S_{-}g+Q_{-}S_{-}g=S_{-}g$. This means: the function
$\Bbb{R}_{+}\ni\lambda\rightarrow S_{\cal K}(\lambda)g(\lambda)$
has an analytic continuation into $\Bbb{C}_{-}$ given by
\[
(S_{-}g)(z)=S_{\cal K}(z)g(z)=\sum_{\zeta\i{\cal R}\cup{\cal P}_{-}}
\frac{S_{-1,\zeta}g(\zeta)}{z-\zeta}-\frac{1}{2\pi i}
\int_{-\infty}^{\infty}\frac{S_{-}(\lambda)g(\lambda)}{\lambda-z}d\lambda,\quad
z\in\Bbb{C}_{-}.
\]
This is true for all $g\in{\cal H}^{2}_{-}$. Putting, for example,
$g(z):=k(z-i)^{-1},\;k\in{\cal K}$ then one obtains (34), where the first term is the
"main part" and
\[
H_{\cal K}(z)=\sum_{\zeta\in{\cal R}\cup{\cal P}_{-}}
\frac{S_{-1,\zeta}}{\zeta-i}-\frac{1}{2\pi i}\int_{-\infty}^{\infty}
\frac{S_{-}(\lambda)}{(\lambda-i)(\lambda-z)}d\lambda
\]
is the holomorphic part of $S_{\cal K}(\cdot)$ in $\Bbb{C}_{-}.\quad\Box$

\vspace{3mm}

Note that, according to Proposition 1, the terms
\[
z\rightarrow\frac{S_{-1,\zeta}g}{z-\zeta_{0}},\quad g\in{\cal K},\,\zeta_{0}
\in{\cal R}
\]
for resonances are "true" Gamov vectors because for simple poles
$\ker L_{+}(\zeta_{0})=\mbox{ima}\,\mbox{Res}_{z=\zeta_{0}}S_{\cal E}(z)=
\mbox{ima}\,S_{-1,\zeta_{0}}$.

\section{References}
\begin{enumerate}
\item Bohm, A.: Quantum Mechanics, Springer Verlag Berlin 1979\\
\item Aigular,J. and Combes, J.M.: A class of analytic perturbations for one-body 
Schr\"odinger Hamiltonians,\\
Comm. Math. Phys. 22, 269-279 (1971)
\item Simon, B.: Resonances in n-body quantum systems with dilatation analytic
potentials and the foundations of time-dependent perturbation theory,\\
Ann. of Math. 97, 247-274 (1973)
\item Hislop, P.D. and Sigal, I.M.: Introduction to Spectral Theory:\\
With Applications to Schr\"odinger Operators, Springer Verlag 1996\\
\item Bohm, A. and Gadella, M.: Dirac Kets, Gamov vectors and Gelfand Triplets,\\
Lecture Notes in Physics 348, Springer Verlag 1989\\
\item Eisenberg, E.,Horwitz, L.P. and Strauss, Y.:\\
The Lax-Phillips Semigroup of the Unstable Quantum System, in :\\
Irreversibility and Causality, Semigroups and Rigged Hilbert Spaces,\\
Lecture Notes in Physics 504, 323-332, Springer Verlag Berlin 1998\\
\item Strauss, Y.: Resonances in the Rigged Hilbert Space and Lax-Phillips 
Scattering Theory,\\
Internat. J. of Theor. Phys. 42, 2285-2317 (2003)\\
\item Agmon, S.: A Perturbation Theory of Resonances,\\
Communications of Pure and Applied Math. 51, 1255-1309 (1998)\\
\item Baumg\"artel, H.: Generalized Eigenvectors for Resonances in the Friedrichs 
Model and Their Associated Gamov vectors,\\
Rev. Math. Phys. 18, 61-78 (2006)\\
\item Kato, T.: Perturbation Theory for Linear Operators, Springer Verlag 1976\\
\item Baumg\"artel, H.: Resonances of Perturbed Selfadjoint Operators and their 
Eigenfunctionals,\\
Math. Nachr. 75, 133-151 (1976)\\
\item Keldysch, M.V.: On eigenvalues and eigenfunctions of classes of
nonselfadjoint equations,\\
Dokl. Akad. Nauk SSSR 77, 11-14 (1951), in russian\\

\item Baumg\"artel, H. and Wollenberg, M.: Mathematical Scattering Theory,\\
Operator Theory: Advances and Applications, Vol. 9, Birkh\"auser Verlag Basel,
Boston, Stuttgart 1983\\
\item Baumg\"artel, H.: Eine Bemerkung zur Theorie der Wellenoperatoren,\\
Math. Nachr. 42, 359-363 (1969)\\
\item Baumg\"artel, H.: Gamov vectors for Resonances: A Lax-Phillips point of view,\\
Internat. J. of Physics, to appear,\\ 
arXiv: math-ph/0407059 (2004)\\
\item Gadella, M.: A rigged Hilbert space of Hardy-class functions:\\
Applications to resonances,\\
J. Math. Phys. 24 (6), 1462-1469 (1983)

\end{enumerate}

\end{document}